\documentclass[%
aip,
amsmath,amssymb,
preprint,%
]{revtex4-2}

\usepackage{graphicx}
\usepackage{dcolumn}
\usepackage{bm}

\usepackage[utf8]{inputenc}
\usepackage[T1]{fontenc}
\usepackage{mathptmx}
\usepackage{etoolbox}

\makeatletter
\def\@email#1#2{%
	\endgroup
	\patchcmd{\titleblock@produce}
	{\frontmatter@RRAPformat}
	{\frontmatter@RRAPformat{\produce@RRAP{*#1\href{mailto:#2}{#2}}}\frontmatter@RRAPformat}
	{}{}
}%
\makeatother

\begin{document}
	
	
	\title{Asymmetric Chiral Coupling in a Topological Resonator}
	\author{Shushu Shi}
        \thanks{Contributed equally to this work}
	\author{Xin Xie}
        \thanks{Contributed equally to this work}
	\author{Sai Yan}
	\affiliation{Beijing National Laboratory for Condensed Matter Physics, Institute of Physics, Chinese Academy of Sciences, Beijing 100190, China}
	\affiliation{CAS Center for Excellence in Topological Quantum Computation and School of Physical Sciences, University of Chinese Academy of Sciences, Beijing 100049, China}
	\author{Jingnan Yang}
	\affiliation{State Key Laboratory for Mesoscopic Physics and Frontiers Science Center for Nano-optoelectronics, School of Physics, Peking University, 100871 Beijing, China}
    \author{Jianchen Dang}
    \author{Shan Xiao}
   	\author{Longlong Yang}
	\author{Danjie Dai}
	\affiliation{Beijing National Laboratory for Condensed Matter Physics, Institute of Physics, Chinese Academy of Sciences, Beijing 100190, China}
	\affiliation{CAS Center for Excellence in Topological Quantum Computation and School of Physical Sciences, University of Chinese Academy of Sciences, Beijing 100049, China}
	\author{Bowen Fu}
	\affiliation{State Key Laboratory for Mesoscopic Physics and Frontiers Science Center for Nano-optoelectronics, School of Physics, Peking University, 100871 Beijing, China}

	\author{Yu Yuan}
	\author{Rui Zhu}
	\affiliation{Beijing National Laboratory for Condensed Matter Physics, Institute of Physics, Chinese Academy of Sciences, Beijing 100190, China}
	\affiliation{CAS Center for Excellence in Topological Quantum Computation and School of Physical Sciences, University of Chinese Academy of Sciences, Beijing 100049, China}
\author{Xiangbin Su}
\author{Hanqing Liu}
\affiliation{State Key Laboratory of Superlattices and Microstructures, Institute of Semiconductors Chinese Academy of Sciences, Beijing 100083, China}

	\author{Zhanchun Zuo}
	\affiliation{Beijing National Laboratory for Condensed Matter Physics, Institute of Physics, Chinese Academy of Sciences, Beijing 100190, China}
	\affiliation{CAS Center for Excellence in Topological Quantum Computation and School of Physical Sciences, University of Chinese Academy of Sciences, Beijing 100049, China}
	\author{Can Wang}
\thanks{Authors to whom correspondence should be addressed: canwang@iphy.ac.cn, xlxu@pku.edu.cn}
	\affiliation{Beijing National Laboratory for Condensed Matter Physics, Institute of Physics, Chinese Academy of Sciences, Beijing 100190, China}
	\affiliation{CAS Center for Excellence in Topological Quantum Computation and School of Physical Sciences, University of Chinese Academy of Sciences, Beijing 100049, China}
	\affiliation{Songshan Lake Materials Laboratory, Dongguan, Guangdong 523808, China}

    \author{Haiqiao Ni}
    \affiliation{State Key Laboratory of Superlattices and Microstructures, Institute of Semiconductors Chinese Academy of Sciences, Beijing 100083, China}
    \author{Zhichuan Niu}
    \affiliation{State Key Laboratory of Superlattices and Microstructures, Institute of Semiconductors Chinese Academy of Sciences, Beijing 100083, China}
	\author{Qihuang Gong}
	\affiliation{State Key Laboratory for Mesoscopic Physics and Frontiers Science Center for Nano-optoelectronics, School of Physics, Peking University, 100871 Beijing, China}
    \author{Xiulai Xu}
    \thanks{Authors to whom correspondence should be addressed: canwang@iphy.ac.cn, xlxu@pku.edu.cn}
	\affiliation{State Key Laboratory for Mesoscopic Physics and Frontiers Science Center for Nano-optoelectronics, School of Physics, Peking University, 100871 Beijing, China}
	\date{\today}
	
	\begin{abstract}
	Chiral light-matter interactions supported by topological edge modes at the interface of valley photonic crystals provide a robust method to implement the unidirectional spin transfer. The valley topological photonic crystals possess a pair of counterpropagating edge modes. The edge modes are robust against the sharp bend of $60^{\circ}$ and $120^{\circ}$, which can form a resonator with whispering gallery modes. Here, we demonstrate the asymmetric emission of chiral coupling from single quantum dots in a topological resonator by tuning the coupling between a quantum emitter and a resonator mode. Under a magnetic field in Faraday configuration, the exciton state from a single quantum dot splits into two  exciton spin states with opposite circularly polarized emissions due to Zeeman effect. Two branches of the quantum dot emissions couple to a resonator mode in different degrees, resulting in an asymmetric chiral emission. Without the demanding of site-control of quantum emitters for chiral quantum optics, an extra degree of freedom to tune the chiral contrast with a topological resonator could be useful for the development of on-chip integrated photonic circuits.
	\end{abstract}
	
	\maketitle
	

    The integrated optical nanostructures with the ability to manipulate light-matter interaction  are important building blocks toward realizing scalable optical quantum networks for quantum-information processing \cite{lodahl2015interfacing,reiserer2015cavity,dietrich2016gaas}. In particular, nanophotonic waveguides with the chiral interface can not only confine photons transversely but also enable the transfer of matter-qubit information when embedded with quantum emitters \cite{lodahl2017chiral}, such as quantum dots (QDs) with spin states as stationary qubits \cite{arakawa2020progress}. Transverse optical confinement in waveguides results in longitudinal and transverse components of electromagnetic fields \cite{bliokh2015transverse,aiello2015transverse}, which leads to polarization-momentum locking. Consequently, when quantum emitters with polarization-dependent transitions couple with the transverse spin angular momentum with the same polarization carried by the waveguide’s electromagnetic fields, the spontaneous emissions radiate unidirectionally as the signature of chirality \cite{lang2022perfect}. The chiral coupling has been demonstrated in many structures \cite{lodahl2017chiral,bliokh2015spin}, including nanofibres \cite{petersen2014chiral,mitsch2014quantum,sayrin2015nanophotonic}, dielectric nanobeam waveguides \cite{coles2016chirality,coles2017path,luxmoore2013optical,luxmoore2013interfacing,javadi2018spin,xiao2021position,xiao2021chiral}, and photonic crystal waveguides \cite{sollner2015deterministic,young2015polarization,mahmoodian2016quantum}. However, the backscattering loss during the propagation of light in those devices is inevitable, which limits their performance in chiral optical circuits.
	
    The recent emergence of topological photonics provides a robust approach to manipulate chiral light-matter interactions \cite{iwamoto2021recent,ozawa2019topological}. The propagation of light in the waveguide formed by topological photonic crystals is immune to deformations and disorders \cite{he2019silicon,shalaev2019robust,yoshimi2021experimental,yoshimi2020slow}. The chiral edge states at the interface of two different topological photonic crystals with the same band gaps have been utilized to realize chiral coupling with handed circular dipoles \cite{barik2018topological,barik2016two,parappurath2020direct}. These edge states in topological waveguides protect the transmission of photons from scattering around sharp bends and defects, which allows the design of nanophotonic resonators possessing chiral spin-photon interfaces \cite{mehrabad2020chiral,jalali2020semiconductor,yang2018topological,ma2019topological,barik2020chiral,xie2021topological}. The Purcell-enhanced strong coupling between a quantum emitter and a topological resonator has been demonstrated with attractive prospects \cite{barik2020chiral}, such as designing photonic integrated circuits for quantum-information processing and studying quantum many-body dynamics \cite{pichler2015quantum}. However, the realization of spin-momentum locking effect in the topological resonator interface mainly depends on the position of quantum emitters inside the resonator \cite{barik2020chiral}, which is challenging to be implemented in photonic integrated circuits.

    Here, we propose an extra degree of freedom to tune the asymmetric emission of chiral coupling in a topological resonator through the magnetic field. The resonator is formed by interfacing two topologically distinct valley photonic crystals (VPCs) with sharp bends of $60^{\circ}$. Single QDs are embedded in the resonator to excite the resonator modes and generate exciton states with opposite circular polarizations when applied with a magnetic field in parallel with QDs’ growth direction. Different unidirectional emissions are achieved as QD exciton states emissions selectively coupled with the counterpropagating edge modes in the interface of the resonator. The chiral coupling is experimentally demonstrated to establish the chirality nature of the resonator. Then, an unequal contrast of the unidirectional emission intensities of opposite directions is simultaneously observed when two branches of QD's emissions are spectral-overlap with a resonator mode in different degrees. The different coupling strength between QD's emissions and a resonator mode can be used to tune the chiral contrast, which is a effective method for tuning chirality in solid systems instead of changing the position of the quantum dots.

	\begin{figure}[b]
	\includegraphics[width=10.6cm]{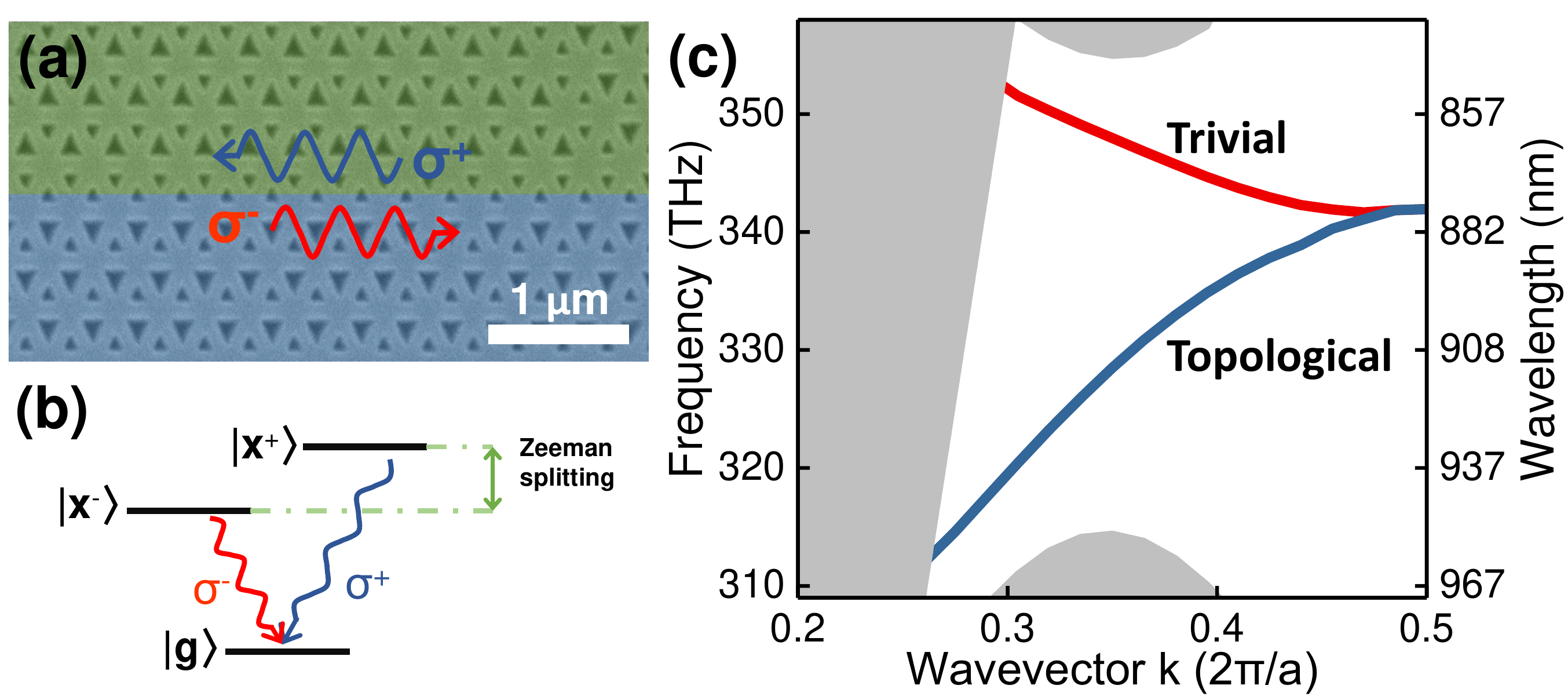}
	\caption{(a) Schematic of the topological chiral interface. (b) Schematic of QD level structure under a magnetic field in Faraday configuration. (c) Dispersion curves for the edge states formed at the bearded interfaces.}
\end{figure}

	Figure 1(a) shows the scanning electron microscope (SEM) image of a valley-Hall topological interface between two topologically distinct VPCs colored in green and blue, respectively. The VPCs are formed by a honeycomb lattice of two inverted equilateral triangular airholes with different side lengths of $L_1$ and $L_2$. As $L_1 \neq L_2$, the spatial inversion symmetry of the VPCs breaks, which causes the formation of a topological band gap. The two types of VPCs share the same side lengths of $L_1$ and $L_2$, resulting in the same band gap. However, the triangular airholes of two VPCs are inverted from each other. The signs of nonzero valley Chern numbers at the $K$ and $K’$ points in two VPCs are opposite, which differ in the topological properties of two VPCs. Therefore, topological edge states are formed at the interface of the two VPCs. The demonstration of chirality in the topological interface requires a pair of circularly polarized dipoles to be selectively coupled to the counterpropagating edge states. QDs can be used as spin-polarized dipole sources and easily integrated in different systems. Figure 1(b) depicts the QD transition under a magnetic field with a Faraday configuration, as the exciton states degenerate due to Zeeman splitting \cite{kuther1998zeeman,wu2020electron,peng2017probing}. Two exciton states are formed with opposite circular polarized emissions \cite{bayer2002fine} of $\sigma^+$ and $\sigma^-$.

    Figure 1(c) shows the dispersion curves of edge states for the interface with simulation parameters of lattice constant a = 255 nm, the side lengths of the two triangular holes $L_1 = 1.3a/\sqrt{3}$ and $L_2=0.7a/\sqrt{3}$. The simulation is performed by the three-dimensional finite-difference time-domain (FDTD) method. The blue line in Fig. 1(c) corresponds to the topological edge state, while the red line indicates the trivial state. The transmission of chiral coupling in the wavelength range of the blue line is topologically protected \cite{yoshimi2020slow}. It is noted that the QD ensemble inhomogeneous emission in our experiment corresponds to the fast-light regime at the bearded interface with a wavelength around 940 nm, as shown in Fig. 1(c).

		\begin{figure}[h]
		\includegraphics[width=10.6cm]{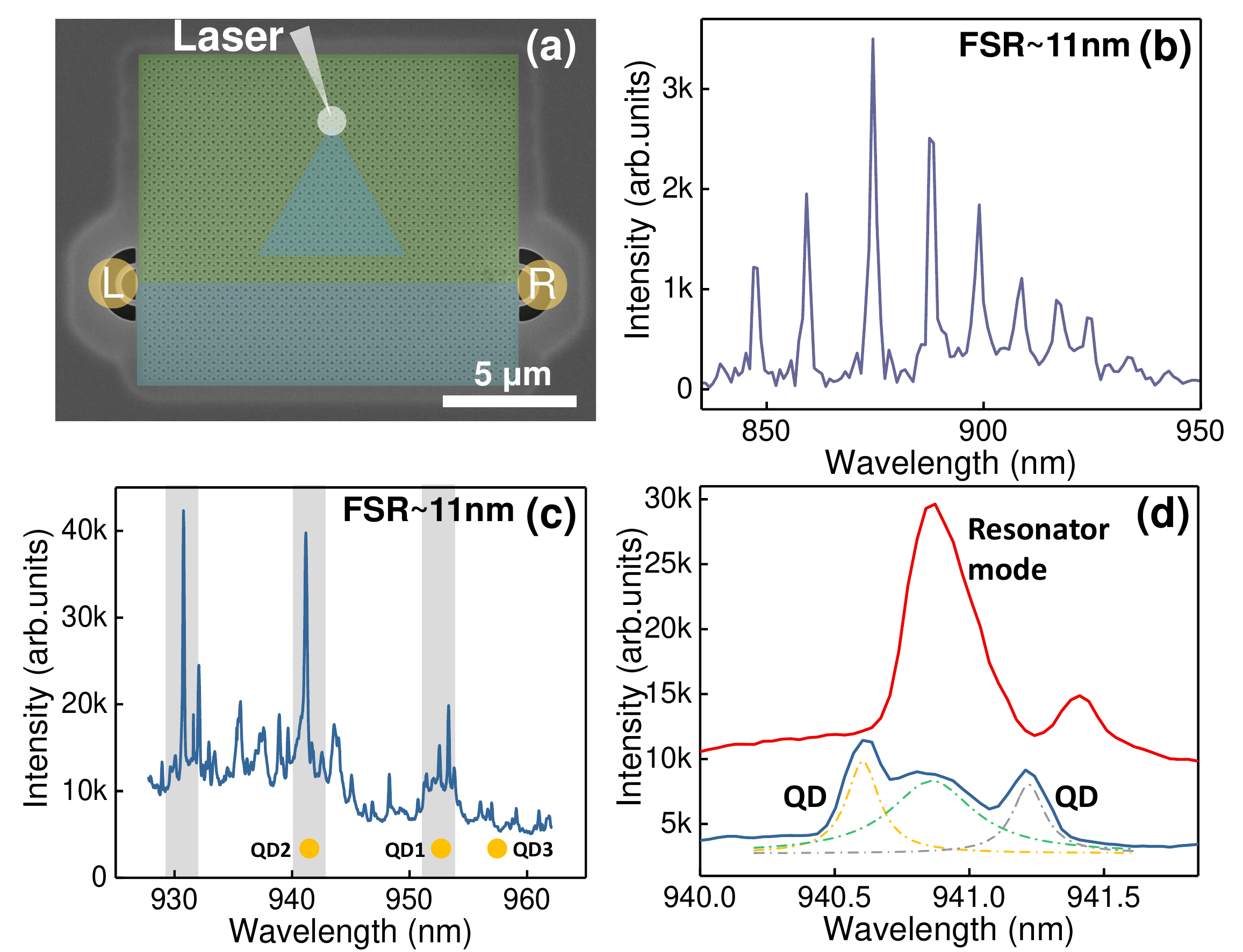}
		\caption{(a) False-color SEM image of a Vally-Hall topological resonator (blue triangle) coupled to a topological waveguide (the interface between green and blue squares) with two grating couplers (yellow). The white circle indicates the excitation spot on the tip of the resonator. (b) Simulated mode distribution of a topological resonator with the averaged FSR of about 11 nm. (c) PL spectrum from a topological resonator embedded with QDs. The gray regions correspond to the distribution of resonator modes. The averaged FSR ($\sim$ 11 nm) from the experimentally measured spectrum matches well with the simulated result. (d) PL spectra of a resonator mode with a Q factor of 2600 under low excitation power (blue line) and high excitation power (red line) at zero magnetic field with two QDs nearby. The half-height width of the mode is 0.358 nm. }
	\end{figure}

	Because of their robustness against sharp bends of $60^{\circ}$ and $120^{\circ}$, the topological edge states can be used to generate a topological resonator, in which the topologically protected chiral coupling with enhanced light–matter interaction is expected \cite{barik2020chiral}. The SEM image of a topological resonator is shown in the inset of Fig. 2(a). The topological resonator is coupled with a topological waveguide, which enables the investigation of chiral coupling between QDs and the edge states inside the resonator. The topological waveguide is terminated by two grating couplers to probe the chiral coupling through a 0.8 numerical aperture objective lens. For fabrication, a single layer of InGaAs QDs is grown in a 150 nm GaAs membrane on a sacrificial layer of AlGaAs. The resonator pattern is transferred on the slab through electron beam lithography and followed by inductively coupled plasma etching. Then, hydrofluoric acid etching is used to remove the exposed AlGaAs sacrificial layer, leaving a suspended membrane with designed structures. The experimental measurements are implemented on a three-axis piezoelectric positioner in a helium bath cryostat under 4.2 K.
	
	The quantum dots on the tip (colored in white circle) of the resonator is nonresonantly excited by a continuous-wave laser of 532 nm. The PL spectrum in Fig. 2(c) collected from the same spot shows the modes of the topological resonator under high excitation power with a free-spectral range (FSR) of about 11 nm. The sharp peaks in Fig. 2(c) are single QDs with half-height widths of around 0.15 nm. Figure 2(b) shows the simulated multiple modes of a topological resonator with the same lattice constant a and the same number of unit cells on the resonator edge as the sample in Fig. 2(a). The average FSR obtained by Lorentz fitting the simulated modes in Fig. 2(b) is about 11 nm. The experimental average FSR matches well with the simulation result. We chose the resonator with a zigzag interface to simulate the distribution of the resonator modes because the exist of slow light modes in the simulation of the resonator with a bearded interface makes it difficult to extract FSR of the whispering gallery modes. The FSR of the whispering gallery modes mainly depends on resonator side length \cite{mehrabad2020chiral}, instead of the types of the interface of the resonator. Figure 2(d) depicts the PL spectrum (red line) under high excitation power, which reveals the resonator mode, along with the PL spectrum (blue line) under low excitation power, which indicates quantum dot emissions close to the resonator mode with a Q factor of 2600. The emission energy of quantum dots can be tuned by a magnetic field \cite{kim2011magnetic} to get into resonance with the resonator mode.

	\begin{figure}[b]
	\includegraphics[width=10.6cm]{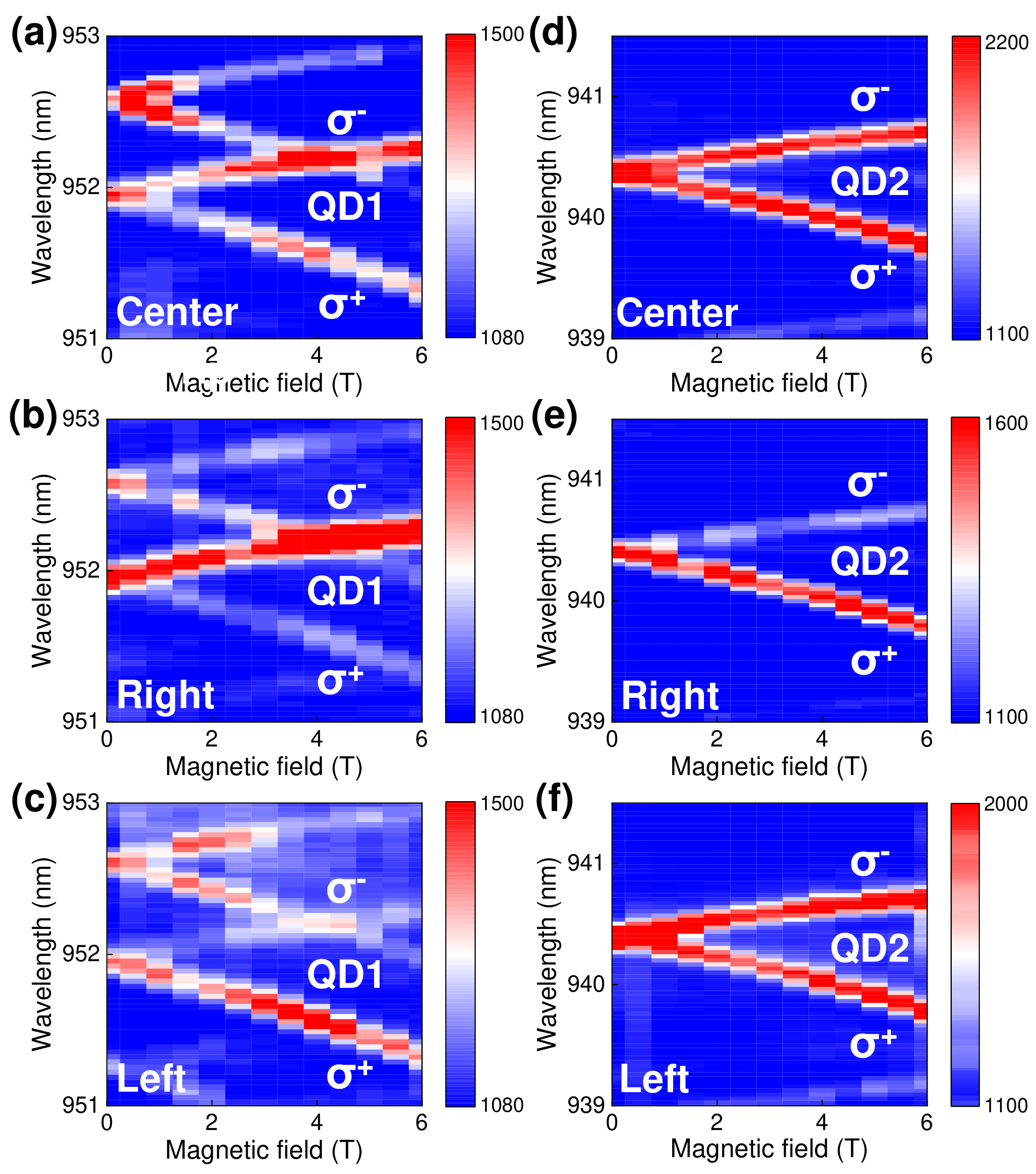}
	\caption{The PL spectra as a function of the magnetic field collected from the excitation spot of (a) QD1 (around 952 nm) and (d) QD2 (around 940.4 nm). The transmission PL spectra  collected from the right grating coupler and left grating coupler from (b)-(c) QD1 and (e)-(f) QD2, respectively.}
\end{figure}

	\begin{figure}[t]
	\includegraphics[width=9.6 cm]{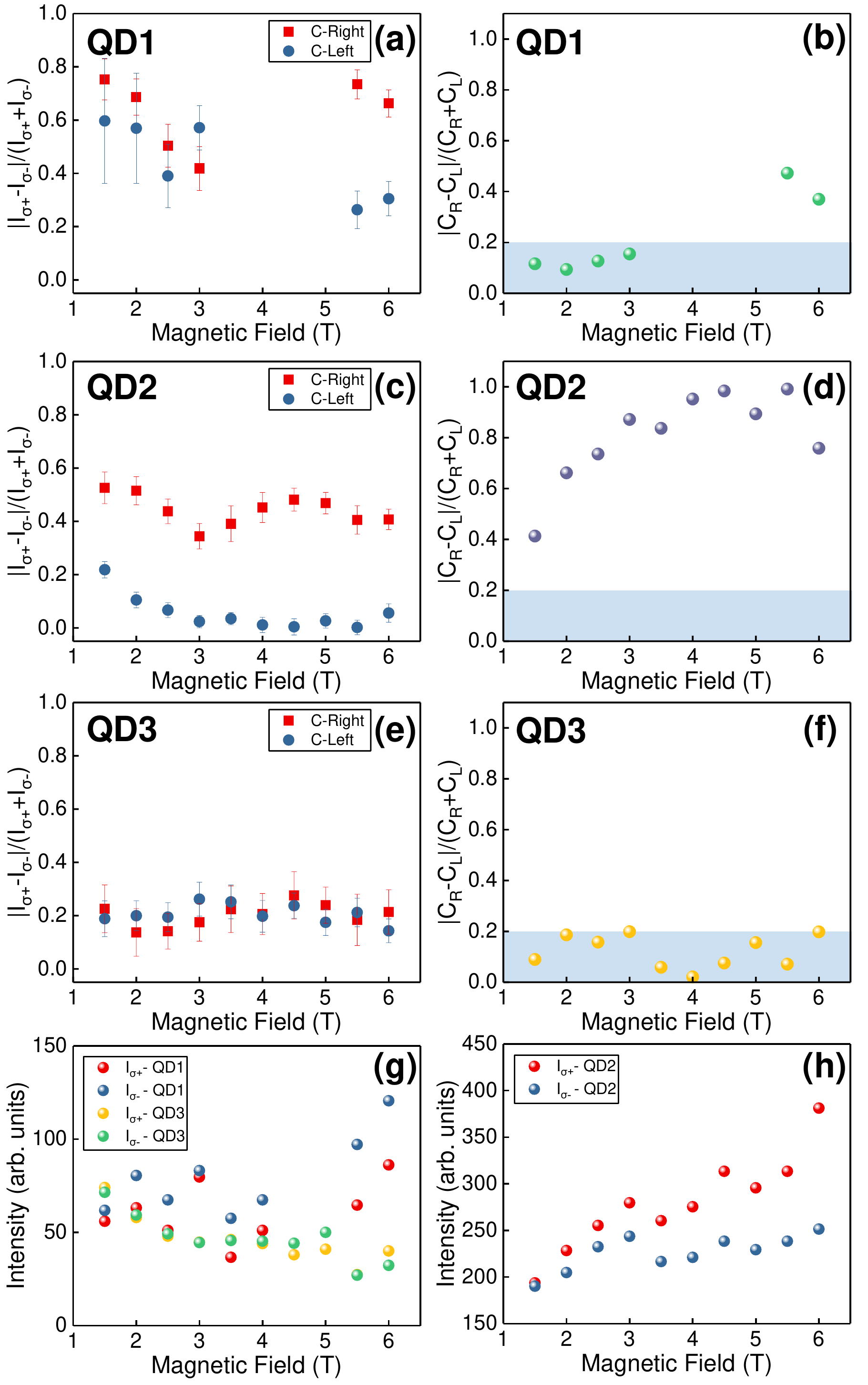}
	\caption{The chiral contrast of (a) QD1, (c) QD2 and (e) QD3 measured from the right (red squares) or left (blue circles) grating coupler as a function of the magnetic field. The contrasts between the corresponding chiral contrasts collected from two grating couplers for (b) QD1, (d) QD2 and (f) QD3 under each magnetic field. The blue region indicates the range of the contrasts in (f) which are under 0.2. The intensity variation of the two spin states from (g) QD1 and QD3, (h) QD2 as a function of the magnetic field during central collection.}
\end{figure}	

	The PL spectra of QD1 and QD2 shown by yellow circles in Fig. 2(c) near the resonator modes are measured as a function of the magnetic field. Under the magnetic field in Faraday configuration, the QD transition occurs due to Zeeman splitting, forming a pair of circularly polarized emissions with opposite spins, which can be spectrally resolved as two branches. The signals of QD1 (around 952 nm) and QD2 (around 940.4 nm) are respectively collected from the excitation spot, left grating coupler and right grating coupler, which are shown in Fig. 3. Both branches of QD1 (around 952 nm) are observed in Fig. 3(a). Only the single branch of QD1 with a circular polarization of $\sigma^-$ is observed in Fig. 3(b), while the $\sigma^+$ branch of QD1 is observed in Fig. 3(c). The transmission spectra exhibit directionality, which is a signature of chiral coupling \cite{xiao2021chiral}. As the topological resonator's chirality is demonstrated, an abnormal transmission result is observed simultaneously in the same spectrum. Figure 3(d) depicts the two branches of QD2 (around 940.4 nm) with measured from the excitation spot. However, the unidirectionality of transmission is more obvious in Fig. 3(e) collected from the right grating coupler comparing to that in Fig. 3(f) collected from the left grating coupler, which indicates the asymmetric chiral coupling for different transmission directions.
	
	To investigate the observed asymmetric chiral coupling in the topological resonator, we calculate the chiral contrasts from each grating coupler as a function of the magnetic field. The chiral contrast C is defined as $C = |I_{\sigma^+}-I_{\sigma^-}|/(I_{\sigma^+}+I_{\sigma^-})$, where $I_{\sigma^-} (I_{\sigma^+})$ denotes the intensity of the upper (lower) branch. The QD3 (around 958 nm) which is away from the resonator modes is also measured for comparison. Figures 4 show the chiral contrasts with the error bar extracted from PL spectra of QD1, QD2 and QD3 collected from left and right grating couplers, respectively. It is noted that the chiral contrast depends on the position of the quantum emitter along the interface of the resonator and the different positions of QD1 and QD2 cause the reversed directional emissions \cite{barik2020chiral}. The chiral contrasts of QD3 measured from left and right grating coupler are almost equal to each other in Fig. 4(e). However, in Fig. 4(a), the difference between the chiral contrasts of QD1 becomes larger under high magnetic field. Moreover, the chiral contrasts of QD2 collected from left and right grating coupler are different under each magnetic field, as shown in Fig. 4(c). The distinction in chiral contrasts indicates that one branch of QD emissions is more affected by the coupling with a resonator mode.

    To further quantify the asymmetric chiral coupling, we define the contrast of two chiral contrasts from two grating couplers with opposite transmission directions as  $C_C = |C_R-C_L|/(C_R+C_L)$, where $C_R (C_L)$ denotes the chiral contrat obtained in the spectrum from right (left) grating coupler. The $C_C$ of QD3 (around 958 nm) under each magnetic field in Fig. 4(f) are ranging from 0 to 0.2 (colored in blue), which can be attributed to the fabrication defects of grating couplers and the measurement error. The $C_C$ of QD1 increases under high magnetic field in Fig. 4(b). Figure 4(g) shows that the intensities of two spin states of QD1 is enhanced under high magnetic field during central collection, while the intensities of two spin states of QD3 are not affected by the resonator mode. And all the $C_C$ of QD2 in Fig. 4(d) are far away from the range (colored in blue) and increase to the maximum at 5.5 T. The rise of $C_C$ is attributed to that the two branches of QD emissions are enhanced by the resonator mode in different degrees, which can be demonstrated by the different intensity enhancement of two spin states of QD2 shown in Fig. 4(h). By introducing the coupling between two spin states of a single QD and a resonator mode, a rise of chiral contrast for one direction and a decrease of that for the other direction are achieved.

	In conclusion, we demonstrate the asymmetric emission of chiral coupling in a topological resonator. A waveguide is coupled with the topological resonator to read out the chiral coupling inside the resonator. The asymmetric chiral coupling is demonstrated by calculating the contrast between chiral contrasts collected from two grating couplers under each magnetic field. When two branches of a QD emissions coupled with a resonator mode in different degrees, the magnetic field modulated asymmetric chiral contrast is observed. It offers an extra degree of freedom to tune the chiral contrast besides changing the position of quantum emitters along the interface of topological resonators. In addition, it offers the ability to enhance the light-matter interaction simultaneously. Such a tunable in-plane asymmetric emission of chiral coupling is beneficial for the development of complex optical circuits on a chip.
	
	\begin{acknowledgments}
		This work was supported by the National Key Research and Development Program of China (Grant No. 2021YFA1400700), the National Natural Science Foundation of China (Grants Nos. 62025507, 11934019, 92250301, 11721404, 62175254 and 12204020), the Strategic Priority Research Program (Grant No. XDB28000000) of the Chinese Academy of Sciences.
	\end{acknowledgments}

%

\end{document}